**Magneto-Optical Spectroscopy of Anatase TiO$_2$ Doped with Co**

Tomoteru FUKUMURA[1,*], Yasuhiro YAMADA[1], Kentaro TAMURA[1,2], Kiyomi NAKAJIMA[3], Toyomi AOYAMA[4], Atsushi TSUKAZAKI[1], Masatomo SUMIYA[5], Shunro FUKE[5], Yusaburo SEGAWA[2], Toyohiro CHIKYOW[3,7], Tetsuya HASEGAWA[6,7], Hideomi KOINUMA[4,7] and Masashi KAWASAKI[1,7]

[1] *Institute for Materials Research, Tohoku University, 2-1-1 Katahira, Aoba-ku, Sendai 980-8577, Japan*

[2] *Photodynamics Research Center, RIKEN, 519-1399 Aoba, Aramaki, Aoba-ku, Sendai 980-0845, Japan*

[3] *National Institute for Materials Research, 1-1 Namiki, Tsukuba 305-0044, Japan*

[4] *Materials and Structures Laboratory, Tokyo Institute of Technology, 4259 Nagatsuta, Midori-ku, Yokohama 226-8503, Japan*

[5] *Department of Electrical and Electronic Engineering, Shizuoka University, 3-5-1 Johoku, Hamamatsu 432-8561, Japan*

[6] *Frontier Collaborative Research Center, Tokyo Institute of Technology, 4259 Nagatsuta, Midori-ku, Yokohama 226-8503, Japan*

[7] *Combinatorial Materials Exploration and Technology, 1-1 Namiki, Tsukuba 305-0044, Japan*

(Received

Magneto-optical spectroscopy of a transparent ferromagnetic semiconductor, anatase TiO$_2$ doped with Co, is carried out at room temperature. A large magneto-optical response with ferromagnetic field dependence is observed throughout from ultraviolet to visible range and increases with increasing Co content or carrier concentration. The magnitude of magnetic circular dichroism (MCD) per unit thickness has a peak around the absorption edge such a huge value of ~10400 degree/cm at 3.57 eV for a 10 mol% Co-doped specimen. Although the results are not sufficient to prove that the ferromagnetism is in the ordinary framework of diluted magnetic semiconductors, the coexistence of Co impurity and mobile carrier is shown to transform the band structure of host $TiO_2$ to generate ferromagnetism.

Wide-bandgap diluted magnetic semiconductors (DMSs) are expected to have higher $T_C$ than the most extensively studied $Ga_{1-x}Mn_xAs$ ($T_C$ ~110 K),[1] which is promising for the realization of future spintronic devices. Matsumoto *et al.* have recently reported that an oxide semiconductor $TiO_2$ can be ferromagnetic above 400 K for both anatase and rutile phases by the doping of Co.[2,3] There are several papers reporting ferromagnetic anatase $TiO_2$ doped with Co.[4-8] Some support Matsumoto's discovery but others claim that the ferromagnetism is caused by the precipitation of Co metal or other compounds.

Generally, the magnetization measurement has been used for the evidence of ferromagnetism. However, careful analysis of the measurement is needed for thin film specimens to rule out the possibilities of the magnetic signal being provided from such impurities in the substrate and precipitation of ferromagnetic compounds.[9] In particular, it is difficult to rule out the existence of ferromagnetic precipitation even when the impurity phases cannot be identified by X-ray diffraction. In order to claim an intrinsic ferromagnetism, ferromagnetic response from the carriers around the band edge interacting with magnetic impurities is a more straightforward probe. To this end, the observation of ferromagnetic magneto-optical response at the bandgap energy of the host semiconductor is quite convincing evidence. Indeed, a magnetic circular dichroism (MCD) study that reported ZnO doped with Co or Ni to be ferromagnetic[10,11] revealed that the characteristic is likely due to ferromagnetic precipitations or is magneto-optically paramagnetic, respectively.[12] Here we report on the MCD study for anatase $TiO_2$ doped with Co, which we call $Ti_{1-x}Co_xO_2$ ($x$: Co content) hereafter.

Epitaxial $Ti_{1-x}Co_xO_2$ thin films were fabricated using a laser molecular beam epitaxy apparatus partly in a combinatorial fashion.[13,14] $LaSrAlO_4$ (001) substrate ($K_2NiF_4$ structure, $a$ = 0.3754 nm, $c$ = 1.263 nm) was used instead of $LaAlO_3$ and SrTiO$_3$ substrates used in the previous studies,[4,15)] because LaAlO$_3$ gives large interference fringes in MCD spectra possibly due to the twin structure and the bandgap of SrTiO$_3$ is similar to that of TiO$_2$ to inhibit us from taking transmission measurements around the bandgap. The ceramic target was ablated by KrF excimer laser pulses at a temperature of 750°C in an oxygen pressure of $1 \times 10^{-6}$ Torr or $5 \times 10^{-7}$ Torr to deposit about 30 nm thick films. The reflection high energy electron diffraction pattern was kept as streaks throughout the film deposition. The X-ray diffraction verified the (001)-oriented anatase phase without any impurity phase. The transmission electron microscope measurements confirm the absence of Co metal precipitation even for the films with $x = 0.1$. The Co content in most of the films was evaluated with electron probe microanalysis, whereas for the films with the most diluted Co content of less than 1 mol%, the Co content was estimated with secondary ion mass spectroscopy. The transmission and reflectance spectra were measured by conventional spectrometer with a normal incidence to deduce the absorption spectra of the films. The electron carrier concentration $n_e$ was evaluated with Hall effect measurement at 300 K. The MCD measurement was carried out with Faraday configuration in a magnetic field parallel to the out-of-plane (001) axis. Alternating right and left circularly polarized monochromatic lights from a Xe lamp (50 kHz) were produced by a photoelastic modulator. The transmitted light was detected by a photomultiplier tube and lock-in amplifier.

    Figure 1(a) shows the absorption spectra for Ti$_{1-x}$Co$_x$O$_2$ ($x = 0, 0.1$) films at 300 K. These films are almost transparent from visible to ultraviolet range. A sharp interband absorption edge appears around 3.5 eV and the in-gap absorption increases with doping Co. Figure 1(b) shows the MCD spectra measured in 1 T at 300 K for Ti$_{1-x}$Co$_x$O$_2$ ($x = 0$, 0.03, 0.1) films. The TiO$_2$ film shows negligible MCD with small paramagnetic

structures above 3.7 eV. These structures presumably originate from the LaSrAlO$_4$ substrate. (See the similar spectrum for the substrate alone with strong temperature dependence in Fig. 1(c) having paramagnetic field dependence.) With increasing $x$, negative MCD around ~2 eV and positive MCD around 3.5-4.5 eV develop. The Ti$_{0.9}$Co$_{0.1}$O$_2$ films with higher $n_e$ show a sharp positive peak at 3.57 eV around the band edge and the magnitude of MCD per unit thickness is as large as ~10400 degree/cm, which is much larger than that of infrared optical isolator materials such as Bi-doped iron garnet (~10$^3$ degree/cm at 1.3 μm). The MCD spectrum of 9 nm-thick Co deposited on TiO$_2$ film is featureless with large magnitude as shown in Fig. 1(c), and is completely different from that of Ti$_{1-x}$Co$_x$O$_2$. Thus, the possibility of bulk precipitation of Co metal in the film can be excluded. Figure 2 shows the magnetic field dependence of the MCD signal for Ti$_{1-x}$Co$_x$O$_2$ ($x$ = 0.03, 0.1) at $h\nu$ = 2.20 eV and 3.57 eV ($h\nu$: photon energy). The magnetic field dependence of the MCD signal clearly shows ferromagnetic behavior. The origin of the variation in coercive force is unclear, as yet.

Figure 3 shows MCD spectra for Ti$_{1-x}$Co$_x$O$_2$ ($x$ = 0, ~0.001) films in 1 T at 300 K. The Ti$_{0.999}$Co$_{0.001}$O$_2$ has a high $n_e$ over 10$^{19}$ cm$^{-3}$ and shows appreciable MCD signal, while TiO$_2$ with similar $n_e$ does not. The MCD signal at 2.20 eV for the Ti$_{0.999}$Co$_{0.001}$O$_2$ shows apparent ferromagnetic response as shown in the inset. The Co content in TiO$_2$ is much diluted far below the percolation limit that the long range ferromagnetic spin ordering cannot be explained without the mobile carrier intermediation between the localized spins. It is noted that such diluted Co content of $x$ ~0.001 leads to the ferromagnetism at 300 K in contrast with the fact that the Curie temperature of Ga$_{1-x}$Mn$_x$As decreases from ~110 K for $x$ = 0.05 to ~30 K for $x$ = 0.015.[16]

We shall compare the present results with Ga$_{1-x}$Mn$_x$As, although the band structures of host compounds are very different. TiO$_2$ is believed to be indirect or direct forbidden transition semiconductor with distorted octahedral coordination, whereas GaAs is direct allowed transition semiconductor with tetrahedral coordination. In case of $Ga_{1-x}Mn_xAs$, paramagnetic MCD responses appear at the energies of critical points ($E_0$ ~1.5 eV, $E_0+\Delta_0$ ~1.9 eV, $E_1$ ~3.0 eV, $E_1+\Delta_1$ ~3.2 eV) of host GaAs for $x = 0$, the MCD responses become large and ferromagnetic at $x = 0.005$, and the spectral shapes are broadened with respect to the photon energy at $x = 0.074$.[17] The absorption spectra in Fig. 1(a) give peaks around 3.9 eV and 4.8 eV, which correspond to the critical points (A and B) observed previously by polarized reflection spectroscopy of a single crystal anatase $TiO_2$.[18] Although the MCD behavior around the critical points for pure $TiO_2$ is not clear due to the signal from the substrate, the MCD around the critical points develops for finite $x$ as seen in Fig. 1(b). In addition, the sharp MCD peak at 3.57 eV for $x = 0.1$ is located near the absorption edge. To summarize, the MCD behavior can be qualitatively correlated with the optical absorption spectra, but is still unclear to use as the conclusive evidence of ferromagnetism in the framework of ordinary DMS. However, only the films with $n_e \geq$ ~$10^{17}$ cm$^{-3}$ showed ferromagnetic MCD among a large number of specimens with varying parameters such as $x$ (0-0.1), oxygen pressure ($10^{-7}$-$10^{-4}$ Torr), and growth temperature (400-800°C). In addition, the appearance of ferromagnetism in the electrically conductive specimen doped with very diluted Co strongly supports the ferromagnetism induced by the interaction between the localized $3d$ spin of Co ion and the mobile $3d$ electron doped in the conduction band of the $TiO_2$ host. The temperature dependence of the resistivity for these films is metallic above 50 K and gives slight upturn without divergence at lower temperatures. The doping of Co results in more degenerate behavior than that of nondoped $TiO_2$. These results may imply that itinerant $3d$ electron carriers are provided not only by crystal defects such as oxygen vacancy in host $TiO_2$ but also by Co ion. This is consistent with the result of band calculation,[19] where Fermi level cuts Co $t_{2g}$ band, and might lead to marginal nature of electronic structure in this compound in comparison with ordinary DMS, since most DMS have carriers in $s$-$p$ hybridized bands. However, we admit that none of our films examined so far gives the anomalous Hall effect seen in $Ga_{1-x}Mn_xAs$. For a better understanding of the origin of ferromagnetism in Co-doped $TiO_2$, further systematic experiments and the examination of the influence of oxygen deficiency are needed.

In conclusion, the MCD spectroscopy of anatase $TiO_2$ doped with Co rules out the apparent precipitation of Co metal as a source of ferromagnetic signal. The results support, but not in a conclusive way, that the coexistence of Co impurity and mobile electron carrier are responsible for ferromagnetism. The strong and broad band MCD signal from ultraviolet to visible range in a transparent substance may be attractive for short wavelength magneto-optical applications.

We thank K. Ando for magneto-optical measurement and critical comments. This work was partly supported by the Ministry of Education, Culture, Sports, Science and Technology, Grant-in-Aid for Creative Scientific Research (14GS0204), the Inamori Foundation, and the Murata Science Foundation.

**Figure captions**

Fig. 1. (a) Absorption spectra for $Ti_{1-x}Co_xO_2$ ($x$ = 0, 0.1) films at 300 K. The electric field is perpendicular to (001). The horizontal axis for $x$ = 0.1 is shifted vertically to the dotted line. The values in parentheses denote the carrier concentration at 300 K. The dashed lines (A and B) correspond to the critical point energy in the band structure of anatase $TiO_2$ (ref. 18) (b) The MCD spectra per unit length for $Ti_{1-x}Co_xO_2$ ($x$ = 0, 0.03, 0.05, 0.1) measured in 1 T at 300 K. The values in parentheses denote the carrier concentration at 300 K. The magnetic field dependences of MCD in Fig. 2 are measured at photon energies denoted by closed triangles. The sample denoted with * was grown in $10^{-6}$ Torr oxygen. The rest of the films ($n_e \geq 10^{19}$ cm$^{-3}$) were grown in $5 \times 10^{-7}$ Torr oxygen. (c) The MCD spectra for Co (9 nm) / $TiO_2$ / $LaSrAlO_4$ at 300 K and $LaSrAlO_4$ substrate at various temperatures in 1 T. Note that the unit of vertical scale is not normalized by the sample thickness.

Fig. 2. (a) The magnetic field dependences of MCD per unit length for $Ti_{1-x}Co_xO_2$ ($x$ = 0.03, 0.1) films (a) at 2.20 eV and (b) at 3.57 eV at 300 K. The values in parentheses denote the carrier concentration at 300 K.

Fig. 3. The MCD spectra per unit length for $Ti_{1-x}Co_xO_2$ ($x$ = 0, ~0.001) in 1 T at 300 K. The spectra are shifted vertically for convenience. The horizontal dotted lines correspond to zero absorption coefficients. The values in parentheses denote the carrier concentration at 300 K. Inset shows the magnetic field dependences of MCD per unit length for $Ti_{0.999}Co_{0.001}O_2$ film at 2.20 eV at 300 K.

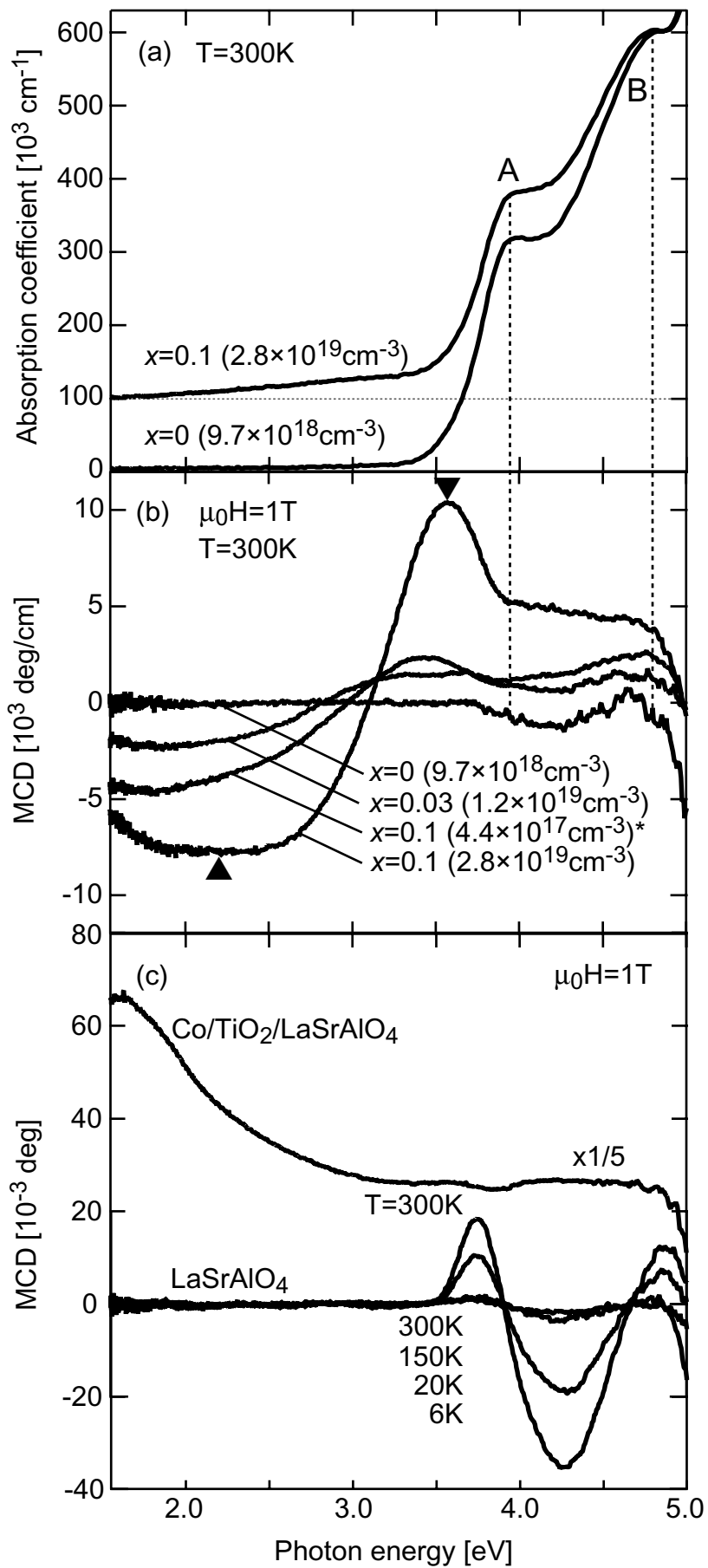

Fig. 1 Fukumura et al. (6.5cm wide)

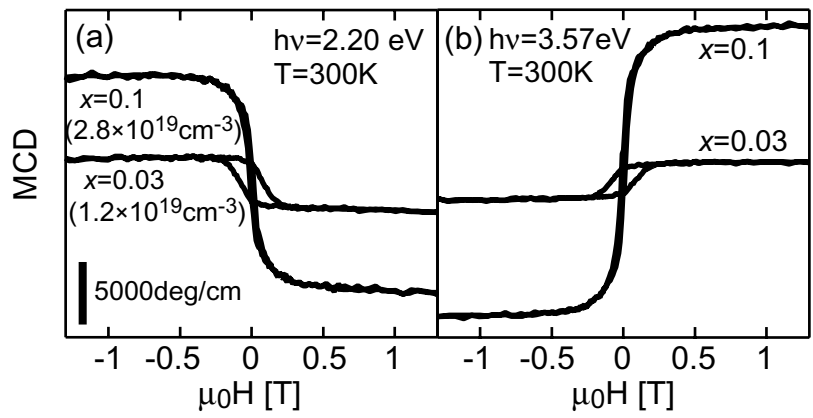

Fig. 2 Fukumura et al. (8cm wide)

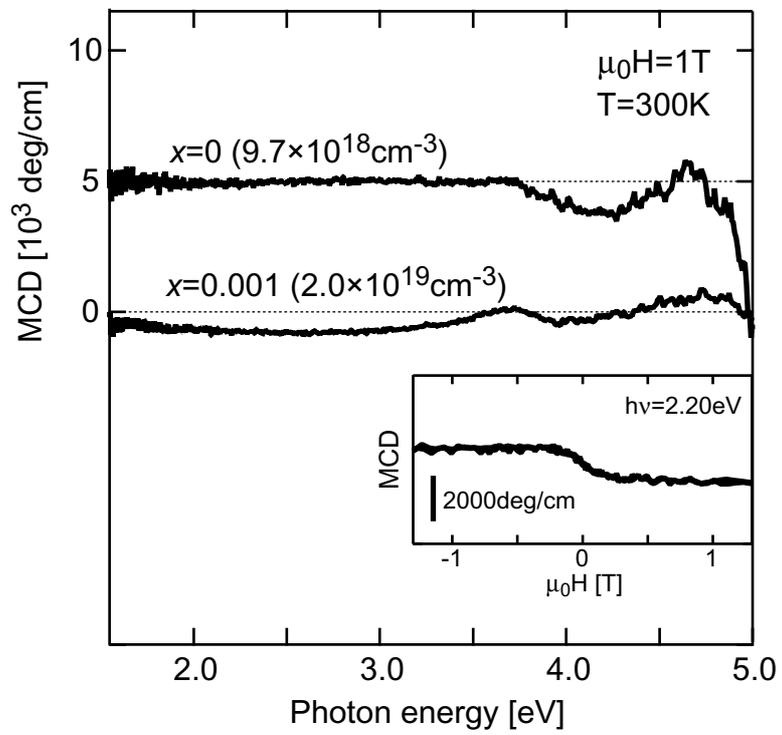

Fig. 3 Fukumura et al. (8cm wide)